\documentclass[runningheads]{llncs}
\usepackage[T1]{fontenc}
\usepackage{graphicx}
\usepackage[table]{xcolor}
\newcolumntype{C}[1]{>{\centering\let\newline\\\arraybackslash\hspace{0pt}}m{#1}}
\usepackage[
	backend=biber,
	style=numeric,
	sorting=nyt,
	giveninits=true,
	maxbibnames=99,
        maxcitenames=2,
	language=british
]{biblatex}

\begin{document}

\title{Multi-Model Ensemble Approach for Accurate Bi-Atrial Segmentation in LGE-MRI of Atrial Fibrillation Patients}
\author{Lucas Beveridge\inst{1} \and Le Zhang\inst{2}}
\authorrunning{L. Beveridge \and L. Zhang}
\institute{School of Computer Science, College of Engineering and Physical Sciences, University of Birmingham, Birmingham, UK \and School of Engineering, College of Engineering and Physical Sciences, \\University of Birmingham, Birmingham, UK}
\date{}

\maketitle

\begin{abstract}
Atrial fibrillation (AF) is the most prevalent form of cardiac arrhythmia and is associated with increased morbidity and mortality. The effectiveness of current clinical interventions for AF is often limited by an incomplete understanding of the atrial anatomical structures that sustain this arrhythmia. Late Gadolinium-Enhanced MRI (LGE-MRI) has emerged as a critical imaging modality for assessing atrial fibrosis and scarring, which are essential markers for predicting the success of ablation procedures in AF patients. The Multi-class Bi-Atrial Segmentation (MBAS) challenge at MICCAI 2024 aims to enhance the segmentation of both left and right atriums and their walls using a comprehensive dataset of 200 multi-center 3D LGE-MRIs, labeled by experts.
This work presents an ensemble approach that integrates multiple machine learning models, including Unet, ResNet, EfficientNet and VGG, to perform automatic bi-atrial segmentation from LGE-MRI data. The ensemble model was evaluated using the Dice Similarity Coefficient (DSC) and 95\% Hausdorff distance (HD95) on the left \& right atrium wall, right atrium cavity, and left atrium cavity. On the internal testing dataset, the model achieved a DSC of $88.41\%$, $98.48\%$, $98.45\%$ and an HD95 of $1.07$, $0.95$, $0.64$ respectively. This demonstrates the effectiveness of the ensemble model in improving segmentation accuracy. The approach  contributes to advancing the understanding of AF and supports the development of more targeted and effective ablation strategies.
\end{abstract}

\keywords{Atrial Fibrillation \and Late Gandolinium-Enhanced MRI \and Multi-class Bi-Atrial Segmentation \and Ensemble Model}

\section{Introduction}
Atrial fibrillation (AF) is the most prevalent form of cardiac arrhythmia, posing significant risks due to its association with increased morbidity and mortality rates. Despite advancements in treatment, the effectiveness of current clinical interventions for AF remain limited, largely due to an incomplete understanding of the atrial anatomical structures that contribute to the perpetuation of AF. Recent developments in medical imaging, particularly Late Gadolinium-Enhanced MRI (LGE-MRI), have facilitated the study of atrial fibrosis and scarring, which are critical markers for predicting the success rates of ablation procedures in AF patients. The extent and distribution of atrial fibrosis, as observed in LGE-MRI scans, offer vital insights into patient-specific treatment planning and the potential for improving clinical outcomes.

At the 27th International Conference Medical Image Computing and Computer Assisted Intervention (MICCAI) 2024, the Multi-class Bi-Atrial Segmentation (MBAS) challenge aims to advance the field of atrial analysis by expanding the scope to include both the left and right atriums and their respective walls \cite{mbas24}. Building on the foundational work of the 2018 Left Atrium Challenge \cite{mbas18}, this competition focuses on leveraging multi-class machine learning models to enhance the segmentation of atrial structures and the identification of key biomarkers, such as atrial volume and fibrosis, from LGE-MRI scans. With the largest dataset of its kind, comprising 200 multi-center 3D LGE-MRIs labeled by experts, the MBAS 2024 challenge offers a robust platform for developing and evaluating artificial intelligence driven approaches to cardiac analysis.

This work presents an ensemble method that combines several machine learning (ML) models to address the challenge of automatic bi-atrial segmentation from LGE-MRI data. Our approach aims to optimize performance on the provided 3-class dataset by leveraging the complementary strengths of these architectures, inspired by the approach of \textcite{ensemble}. We evaluate our ensemble model using the Dice Similarity Coefficient (DSC) and the 95\% Hausdorff distance (HD95), reflecting the assessment of the MBAS challenge. By improving the segmentation of atrial structures, our method contributes to the broader goal of refining ablation strategies and enhancing patient-specific treatment plans for atrial fibrillation.
\section{Methods}
\subsection{Dataset}
The data used to train the ML models was sourced from the MBAS challenge organizers. It consists of 200 multi-center 3D LGE-MRIs, each scan labelled by three experts. The MBAS challenge organizers themselves split these 200 scans into 70 training, 30 validation and 100 testing. The validation and testing sets were used to externally assess the performance of participants' submissions, and so the only labelled data available to participants were the 70 training scans.
These scans had dimensions of either $40\times640\times640$ or $40\times576\times576$, and consisted of three classes (or four if the background is included as a class): left \& right atrium wall, right atrium cavity, and left atrium cavity.

The 70 labelled training scans were subsequently split into internal training, validation and testing datasets in a ratio of 84:8:8. Each 3D scan was treated as a series of 2D images stacked in the depth dimension, giving a total of 3080 2D images either of size $640\times640$ or $576\times576$. This gave 2587 training images, 246 validation images, and 247 testing images. In line with the approach of \textcite{ensemble}, three different splits were created; the testing dataset was fixed while the training and validation datasets were different for each split. These splits were created randomly across the entire domain of 2D images. This meant that slices of a single 3D scan could potentially be split across training, validation and testing datasets, with the goal of improving generalization.

Regularization was also introduced in the form of data augmentation. This was performed at runtime as a model was being trained. The augmentations applied were horizontal flip, vertical flip, and rotation by an angle, randomly chosen as one of $45^{\circ}$, $135^{\circ}$, $225^{\circ}$, or $315^{\circ}$. Each augmentation had a 10\% probability of being applied to an image, and only one augmentation would be applied to an image.

\subsection{Model Architectures}
Like \textcite{ensemble}, four CNN architectures were used: UNet \cite{unet}, ResNet \cite{resnet}, EfficientNet \cite{effnet}, and VGG \cite{vgg}. All models take a single-channel 2D LGE-MRI slice as input, and output a four-channel image of the same size (either $640\times640$ or $576\times576$). Each output channel represents the logits of each pixel belonging to a particular class: background, wall, right atrium and left atrium respectively. The final activation layer is an $argmax$, converting the four-channel logits to a single-channel integer label. All architectures apart from VGG included regularization in the form of dropout \cite{dropout} at a probability of 20\%.

\subsubsection{UNet}
UNets were first proposed by \textcite{unet} for biomedical image segmentation. They consist of an encoder-decoder architecture with skip connections in-between. The architecture used in this case comprised four downsampling layers, a bottleneck, and four upsampling layers. Each downsampling layer consisted of a convolution with kernel size $4\times4$, stride 2 and padding 1, batch normalization \cite{batch-norm}, dropout and a LeakyReLU activation with negative slope $0.2$. Each upsampling layer mirrored the corresponding downsampling layer, but replaced the convolution with a transposed convolution of the same parameters.

\subsubsection{ResNet}
ResNets were introduced by \textcite{resnet} to address the vanishing gradient problem and facilitate the training of very deep neural networks. As the standard ResNet architecture is designed for image classification, it must be adapted for segmentation. This was done by using a ResNet34 \cite{resnet} as the encoder of a UNet. The decoder consists of upsampling layers followed by convolutions of size $3\times3$, batch normalisation, dropout, and ReLU activations.

\subsubsection{EfficientNet}
EfficientNets were introduced by \textcite{effnet}, utilizing a compound scaling method that uniformly scales all dimensions of depth, width, and resolution, making it computationally efficient while maintaining high performance. EfficientNetB0 was used as the encoder of a UNet, with the same decoder architecture used for the ResNet.

\subsubsection{VGG}
VGG is a classification architecture first proposed by \textcite{vgg}, with a deep but simple convolutional structure consisting of small $3\times3$ kernels. As with the ResNet and EfficientNet, the VGG architecture used in this approach integrates a VGG16 encoder into a UNet for segmentation. The implementation follows that of \textcite{vgg-unet}, where the decoder consists of transposed convolutions of size $2\times2$ and stride 2, followed by regular convolutions, batch normalisation, and ReLU activation.

\subsection{Training and Testing}
Each architecture (UNet, ResNet, EfficientNet and VGG) was trained on each of the three splits, giving a total of 12 models. Each model was trained for 300 epochs using the Adam optimizer \cite{adam} and a OneCycle learning rate scheduler \cite{onecycle}. The initial learning rate was $1\times10^{-4}$, which increased to a maximum of $1\times10^{-2}$, and then annealed to a minimum of $1\times10^{-6}$. A joint loss function comprising cross entropy and the Dice metric was used.

Once all models were trained, the Ensemble model was created and trained for 30 epochs with the same optimizer and learning rate scheduler as above. The loss function was a weighted combination of cross entropy, Dice metric, and log HD95 in the ratio $0.2:0.4:0.4$. The ensemble model wrapped all sub-models with four $1\times1$ convolutions, one for each class. This is effectively a per-class weighted combination of each of the sub-models, and so optimizing the Ensemble is equivalent to finding the weight for each sub-model that gives the best Dice and HD95 metrics per class. The final activation of the Ensemble model is $argmax$, giving a single-channel, integer label. All models were evaluated on the internal testing dataset, comparing their outputs to corresponding ground truth labels. Dice Score and HD95 were used as evaluation metrics.

All models were implemented and trained in PyTorch \cite{pytorch}, using elements from the Medical Open Network for AI (MONAI) \cite{monai}.

\section{Results}
Table \ref{tab:scores} shows the Dice and HD95 scores for every model on each class. In general, the UNet achieves the worst results, consistently achieving the lowest Dice scores and highest HD95 distances in all three classes. This is most likely due to the UNet having the smallest number of parameters amongst all models, therefore forming the simplest model of the LGE-MRI data and its heart atria. This reasoning is reinforced by the EfficientNet and ResNet achieving better results than the UNet; both have deeper architectures and are therefore able to capture the data distribution more accurately. ResNet attains narrowly superior results over EfficientNet, which could be due to its smaller model size \cite{effnet}. However, the difference in scores is small enough that it could instead be due to minor differences such as random weight initialization. Surprisingly, VGG achieves the greatest results of all the sub-models. This is unexpected as the VGG architecture predates ResNets and EfficientNets; both were proposed as an advancement in training speed and accuracy \cite{resnet, vgg, effnet}. This could be because VGGs have a higher complexity than ResNets \cite{resnet}, allowing them to adapt to the complicated problem of multi-class segmentation more easily.

\begin{figure}
    \includegraphics[width=1\textwidth,  trim={4.4cm 7cm 3.5cm 7cm}, clip]{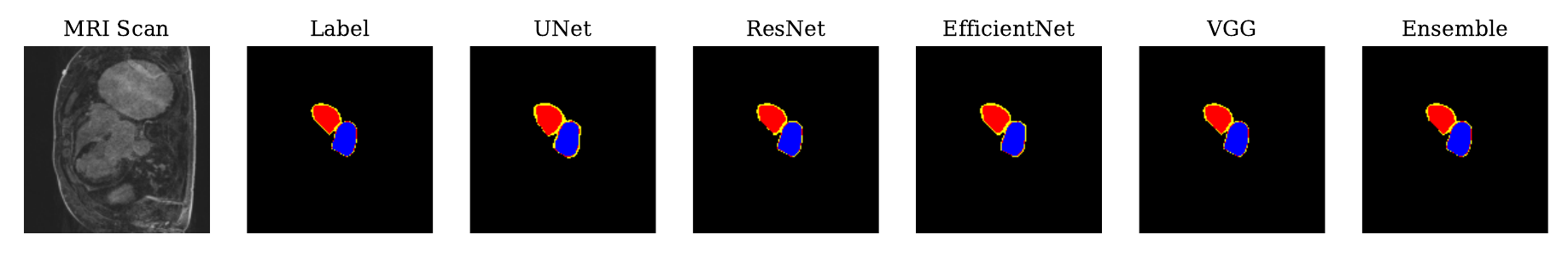}
    \includegraphics[width=1\textwidth,  trim={4.4cm 7cm 3.5cm 7.8cm}, clip]{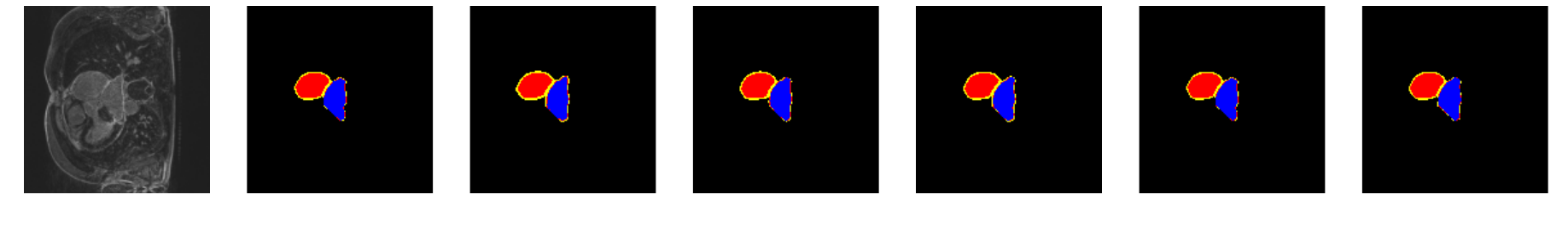}
    \includegraphics[width=1\textwidth,  trim={4.4cm 7cm 3.5cm 7.8cm}, clip]{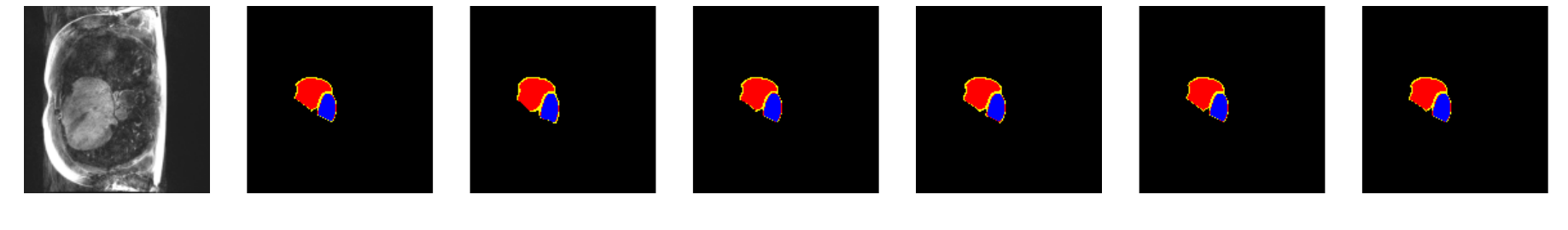}
    \caption{Outputs of each model (UNet, ResNet, EfficientNet, VGG and Ensemble) for three LGE-MRI inputs (MRI Scan), compared to their respective ground truths (Label). The models UNet, ResNet, EfficientNet and VGG were all trained on dataset split A.}
    \label{fig:outputs}
\end{figure}

The Ensemble model achieves the best results, with the highest Dice score and HD95 distance in each class by a considerable amount. This reflects the strength of a model ensemble, able to combine the strengths of multiple, individual models to attain superior results. Figure \ref{fig:outputs} shows the outputs of each model on three MRI scans from the internal testing dataset, along with their ground truth labels. It shows that each model has small inaccuracies, such as the atrial wall being too thick, or the edges of each atrium not being the right shape, and that the Ensemble is able to mitigate these inaccuracies through collective strength.

\begin{table}
\renewcommand{\arraystretch}{1.1}
\definecolor{lb}{RGB}{221, 235, 247}
\definecolor{db}{RGB}{189, 215, 238}
\definecolor{lg}{RGB}{226, 239, 218}
\definecolor{dg}{RGB}{198, 224, 180}
\definecolor{ly}{RGB}{255, 242, 204}
\definecolor{dy}{RGB}{255, 230, 153}
\definecolor{s}{gray}{0.92}
    \centering
    \caption{Dice and HD95 scores for the left \& right atrium wall (Wall), right atrium (RA) and left atrium (LA) of all twelve models - four architectures (UNet, ResNet, EfficientNet and VGG) and three dataset splits (A, B and C). The best Dice and HD95 score for each class is in bold.}
    \begin{tabular}{| C{2.5cm} | C{1cm} | >{\columncolor{lb}}C{1.2cm} | >{\columncolor{lb}}C{1.2cm} | >{\columncolor{lg}}C{1.2cm} | >{\columncolor{lg}}C{1.2cm} | >{\columncolor{ly}}C{1.2cm} | >{\columncolor{ly}}C{1.2cm} |}
        \hline
        & & \multicolumn{2}{>{\columncolor{lb}}c|}{\textbf{Wall}} & \multicolumn{2}{>{\columncolor{lg}}c|}{\textbf{RA}} & \multicolumn{2}{>{\columncolor{ly}}c|}{\textbf{LA}} \\
        \cline{3-8}
        \textbf{Architecture} & \textbf{Split} & Dice & HD95 & Dice & HD95 & Dice & HD95 \\
        \hline
             & A & 64.36 & 8.18 & 89.53 & 6.75 & 91.61 & 4.77 
             \\
        \rowcolor{s}
        UNet & B & \cellcolor{db}65.33 & \cellcolor{db}8.34 & \cellcolor{dg}89.96 & \cellcolor{dg}11.80 & \cellcolor{dy}91.97 & \cellcolor{dy}4.33 \\
             & C & 65.20 &  6.92 & 91.24 & 4.58 & 91.21 & 3.75 \\
        \hline
               & A & 74.88 & 3.92 & 93.80 & 2.32 & 94.89 & 2.40 \\
        \rowcolor{s}
        ResNet & B & \cellcolor{db}73.45 &  \cellcolor{db}4.07 & \cellcolor{dg}94.11 & \cellcolor{dg}2.73 & \cellcolor{dy}94.05 & \cellcolor{dy}2.94 \\
               & C & 74.80 & 3.08 & 95.02 & 2.98 & 94.15 & 1.94 \\
        \hline
                     & A & 72.18 & 4.52 & 93.80 & 3.25 & 94.05 & 2.40 \\
        \rowcolor{s}
        EfficientNet & B & \cellcolor{db}70.96 & \cellcolor{db}3.80 & \cellcolor{dg}94.69 & \cellcolor{dg}2.39 & \cellcolor{dy}93.70 & \cellcolor{dy}2.83 \\
                     & C & 71.99 & 3.30 & 94.75 & 2.67 & 93.47 & 2.65 \\
        \hline
            & A & 75.27 & 3.61 & 93.11 & 2.42 & 94.45 & 2.50 \\
        \rowcolor{s}
        VGG & B & \cellcolor{db}75.31 & \cellcolor{db}3.22 & \cellcolor{dg}94.37 & \cellcolor{dg}2.38 & \cellcolor{dy}92.88 & \cellcolor{dy}2.47 \\
            & C & 75.16 & 3.05 & 94.39 & 2.76 & 94.05 & 2.11 \\
        \hline
        Ensemble & N/A & \textbf{88.41} & \textbf{1.07} & \textbf{98.48} & \textbf{0.95} & \textbf{98.45} & \textbf{0.64} \\
        \hline
    \end{tabular}
    \label{tab:scores}
\end{table}

\section{Discussion and Conclusion}
The results of this study demonstrate the effectiveness of the ensemble approach in multi-class bi-atrial segmentation from LGE-MRI data, with superior performance compared to individual models such as UNet, ResNet, EfficientNet, and VGG. The ensemble model, leveraging the complementary strengths of each architecture, achieved the highest Dice and HD95 scores across all atrial regions, significantly outperforming any single model. This reflects the power of ensemble learning, where model diversity helps to reduce errors and improve generalization. The unexpected performance of VGG, outperforming more modern architectures like ResNet and EfficientNet, suggests that model complexity may play a role in better capturing the intricate features of LGE-MRI data, although further investigation into specific factors like initialization and training dynamics could provide additional insights.

Ultimately, this ensemble-based method offers a robust solution for automatic multi-class segmentation of the left and right atrium from LGE-MRI data, achieving high accuracy and precision as evidenced by the Dice and HD95 metrics. The success of the ensemble model in this context suggests it could be an important tool for advancing atrial fibrillation treatment strategies by providing more reliable segmentation of atrial structures. Future work could explore further refinements, such as hyperparameter tuning or integration of other architectures, to push the boundaries of segmentation accuracy even further.

\printbibliography

\end{document}